\documentclass[preprint]{revtex4}

\usepackage{graphicx}

\begin{document}

\title{Ultrasensitive force and displacement detection using trapped ions}

\author{Michael J. Biercuk}
\altaffiliation{\emph{Present address} School of Physics, University of Sydney, NSW 2006, Australia}
\author{Hermann Uys}
\altaffiliation{\emph{Present address} Council for Scientific and Industrial Research, Pretoria, South Africa}
\author{Joe W. Britton}
\author{Aaron P. VanDevender}
\author{John J. Bollinger}
\email[To whom correspondence should be addressed: ]{michael.biercuk@sydney.edu.au}
\affiliation{NIST Time and Frequency Division, Boulder, CO, 80305}

\date{\today}

\begin{abstract} 
The ability to detect extremely small forces and nanoscale displacements is vital for disciplines such as precision spin-resonance imaging~\cite{Rugar2004}, microscopy~\cite{AFM1986}, and tests of fundamental physical phenomena~\cite{Casimir1998, Friction1999, Gravity1998}.  Current force-detection sensitivity limits have surpassed 1 $\textrm{aN}/\sqrt{\textrm{Hz}}$~\cite{Rugar2001, Lehnert2009} (atto $=10^{-18}$) through coupling of nanomechanical resonators to a variety of physical readout systems~\cite{Cleland2003, Schwab2004, Rugar2004, Lehnert2008, Lehnert2009}.  Here we demonstrate that crystals of trapped atomic ions~\cite{brel88, BiercukQIC2009} behave as nanoscale mechanical oscillators and may form the core of exquisitely sensitive force and displacement detectors.   We report the detection of forces with a sensitivity 390$\pm150$~$\textrm{yN}/\sqrt{\textrm{Hz}}$ (more than three orders of magnitude better than existing reports using nanofabricated devices~\cite{Lehnert2009}), and discriminate ion displacements $\sim$18 nm.  Our technique is based on the excitation of tunable normal motional modes in an ion trap~\cite{Heinzen1990} and detection via phase-coherent Doppler velocimetry~\cite{Berkeland1998, mitchellDoppler}, and should ultimately permit force detection with sensitivity better than 1~$\textrm{yN}/\sqrt{\textrm{Hz}}$~\cite{Maiwald2009}.  Trapped-ion-based sensors could permit scientists to explore new regimes in materials science where augmented force, field, and displacement sensitivity may be traded against reduced spatial resolution.

\end{abstract}

\maketitle

\indent Trapped atomic ions exhibit well characterized and broadly tunable (kHz to MHz) normal motional modes in their confining potential \cite{Bible, Maiwald2009}.  The presence of these modes, the light mass of atomic ions, and the strong coupling of charged particles to external fields makes trapped ions excellent detectors of small forces with tunable spectral response~\cite{Heinzen1990}.  Another advantage is that readout is achieved through resonant-fluorescence detection using only a single laser.  Previous studies have suggested that by using ions it is possible to measure forces approaching the yoctonewton scale, for instance, through experiments on motional heating in Paul traps due to fluctuating electric fields~ \cite{Turchette2000, Deslauriers2004, Labaziewicz2008}, or resonant excitation techniques~\cite{Bible, Drewsen2004}.  In particular, small forces applied to ions in weak trapping potentials (trapping frequencies $\sim$0.1 MHz or lower) can excite micron-scale motional excursions resolvable using real-space imaging~\cite{Drewsen2004, Staanum}.

\indent While the intrinsic sensitivity of trapped ions to external forces and fields is well supported, it remains an experimental challenge to determine the maximum \emph{achievable} sensitivity to a given external excitation as set by systematic limitations including the efficiency of a measurement procedure.  Establishing ions as components in ultrasensitive detectors requires two primary issues to be addressed: a known excitation must be applied to allow precise calibration of the system's response; and it must be possible to compare the results of these experiments with the existing literature on detectors based on integrated nanostructures.   Our aims are to unify the seemingly disparate fields of nanotechnology and atomic devices, through use of comparable experimental conditions and a demonstration of the potential utility of ion-based sensors in nanomaterials characterization.
\\
\indent Our system is a crystal of $^{9}$Be$^{+}$ ions in a Penning trap~\cite{brel88, Major2005, BiercukQIC2009}.  Ions scatter ultraviolet laser light near 313 nm, which is nearly resonant with an internal atomic transition, cooling the ions and providing a direct imaging mechanism (see \textit{Supplementary Information} for details).  Scattered fluorescence is detected using an imaging system connected to a CCD or photomultiplier tube.  For these experiments we focus on ion crystals~\cite{mitchell98, huap98a, jenm05} with $n$$\approx$100 ions in a two-dimensional planar array (perpendicular to the direction of the detection laser), having a diameter $\sim$300 $\mu$m.  
\\
\indent In our detection technique, known as laser Doppler velocimetry~\cite{mitchellDoppler, Berkeland1998}, uniform ion motion at the axial center-of-mass (COM) frequency (set at $\omega_{Z}/2\pi=$ 867 kHz for this work) parallel to the propagation direction of the cooling laser beam modulates the intensity of resonant fluorescence due to Doppler shifts.  Under oscillatory ion motion, and with the laser detuned from resonance near the Doppler-cooling-efficiency maximum~\cite{mitchellDoppler} the intensity of ion fluorescence is modulated at $\omega_{Z}$ with amplitude approximately proportional to the motional amplitude (Fig. 1a).  For uniform light collection from all ions and detection laser power below saturation, the total detection rate of scattered photons is $n\rho\left[1+(2/\gamma) k \dot{z}_{COM}\right]$, where $n$ is the ion number, $\rho$ captures all hardware parameters including quantum efficiency of the detector, the intensity of illumination, etc., $k$ is the wavevector of the Doppler detection laser parallel to the ion motion, and $\dot{z}_{COM}$ is the velocity of the axial center-of-mass coordinate.  %%Assuming steady-state ion oscillation with amplitude $z_{COM}$ and small excitations, we expect a modulation of the total photon scatter rate at frequency $\omega_{COM}$, with fractional amplitude $2\omega_{COM}kz_{COM}/\gamma$.
\\
\indent We consider an impulse-style measurement in which a force, $F_{d}\sin(\omega_{d}t)$ is applied to the ions for a fixed drive time, $t_{d}\gg2\pi/\omega_{d}$, during which the Doppler detection laser is turned off.  As such, radiation damping during the excitation does not occur.  We define $F_{d}$ to be the amplitude of the total force applied to a crystal of $n$ ions, and $F_{d}^{(ion)}$ to be the amplitude of the force applied to a single ion of mass $m$.  After application of this drive pulse, an ion crystal of $n$ ions will undergo a steady-state sinusoidal oscillation with velocity
\begin{equation}
\dot{z}_{COM}(t)=v\sin\left[\omega_{z}t+\phi\right],
\end{equation}

\noindent where for $\left |\omega_{z}-\omega_{d}\right |/\omega_{z}\ll 1$, the amplitude $v$ is given by
\begin{equation}
v=\frac{2F_{d}\omega_{d}}{nm(\omega_{z}^{2}-\omega_{d}^{2})}\sin\left[\frac{(\omega_{z}-\omega_{d})t_{d}}{2}\right],
\end{equation}

\noindent and the oscillation phase $\phi$ is
\begin{equation}
\phi=\frac{(\omega_{d}-\omega_{z})t_{d}}{2},
\end{equation}

\noindent assuming no damping in a harmonic confining potential, $m\omega_{z}^{2}z_{COM}^{2}/2 $.  Near resonance ($\omega_{Z}\approx\omega_{d}$), Eq.~2 can be written $v=(F_{d}/2nm)t_{d}$.  
\\
\indent Ion motion induced by the application of this driving force will be superimposed on a noisy background due to the finite temperature of the axial mode and stray electric fields. The ability to discern a small oscillating signal of known frequency in the presence of a large background is well established through use of phase-sensitive detection~\cite{HorowitzHill}.  In these techniques, shown to be applicable in almost any setting, a signal of interest is discriminated through synchronization of a target system's response to a master oscillator that produces or modulates the excitation.  In doing so, broadband environmental noise is excluded, and only the integrated noise over the narrow bandwidth of the measurement is germane.
\\
\indent The temporal modulation of ion fluorescence due to the applied drive is detected phase-coherently by recording scattered-photon arrival times using a photomultiplier tube relative to a trigger synchronized to the external drive force.  Photon arrival times relative to the drive force are then determined using a time-to-amplitude converter (TAC) or multichannel scaler over $N$ iterations of the experiment.   Any noise, however, is not phased with the drive, and averaging $N$-times increases the contrast of the desired signal relative to background noise by a factor of $\sqrt{N}$.  
\\
\indent Precise characterization of the the achievable force- and displacement-detection sensitivity requires the application of a well calibrated drive force.  Such a force is generated using the electric field from an RF voltage applied to an endcap electrode on the trap, similar to resonant electric-field excitation techniques used in cantilever experiments~\cite{Rugar1997} (Fig. 1b, c).  The electric field at the location of the ions is calibrated using a measurement of the static deflection of a planar ion crystal by sideview imaging under application of a static voltage to the same endcap electrode (see \textit{Supplementary Information}).  As an example, for an applied zero-to-peak voltage of 165$\pm$10 $\mu$V (nominal RF power, $P_{RF}=$ -70 dBm), we determine an electric field of 1.8$\pm$0.1 mV/m at the location of the ions, and a corresponding force $F_{0}^{(ion)}=$ 290$\pm$18 yN per ion. 
\\
\indent  Figure 1d shows a typical histogram of the arrival times of the first detected photon synchronized to the drive for $\omega_{d}=\omega_{Z}$.  For short times no photons are detected due to hardware delays (mainly the response of the acousto-optic modulator (AOM) switch).  Once photon-detection events begin accumulating after $\sim$4 $\mu$s, we find a bunching of photon arrival times  with a period commensurate with the 1.15 $\mu$s period of the COM oscillation.  Phase information is captured in the absolute locations of histogram maxima along the time axis.  This approach allows the target system to have a variable shot-to-shot phase difference from the master oscillator as experimental parameters are swept, thus providing more information than standard lock-in detection in which the detection phase offset is fixed.  
\\
\indent We study the Doppler velocimetry signal as a function of $\omega_{d}$ and $t_{d}$.  The measurement scans the drive frequency and records a histogram of stop-pulse delays relative to the drive-force trigger (the start pulse).  The modulation of the scattered photon rate is plotted as a colorscale, after correcting for an exponential decay factor due to the triggering technique (see \textit{Supplementary Information}).  The first column of Figure 2 (Fig. 2a, d, g, j) shows experimental measurements of COM-mode excitation, while the second column (Fig. 2b, e, h, k) shows theoretical calculations based on the formulae above.  In these calculations we fix $t_{d}$ and $\omega_{Z}$ and allow a variable drive strength and a background, both of which are held the same for all $t_{d}$.  
\\
\indent Theoretical calculations match well with experimental data, replicating both qualitative and quantitative features.  Doppler velocimetry indicates a linear phase shift in the oscillator as $\omega_{d}$ is tuned through resonance, and the excitation of the oscillator is zero when $|\omega_{d}-\omega_{z}|=2\pi/t_{d}$, such that during the driving period the external force and system response desynchronize and resynchronize~\cite{Leibfried2003}.  Moreover, in addition to the central resonance feature, oscillation sidelobes appear, separated by the detuning period $2\pi/t_{d}$, and we have seen ten sidelobes for strong RF excitation.  In the experimental data we also observe a damping of the oscillation strength as a function of the delay time due to radiation pressure from the detection laser.  This effect is not accounted for in the calculations shown in the second column of Fig. 2.  
\\
\indent The linewidth of the resonance scales as the inverse of $t_{d}$ (Fig. 2c, f, i, l).  Here we plot the oscillation amplitude (using the standard deviation of the photon-arrival-time histogram for all $t$ as a proxy), as a function of $\omega_{d}$, and find that the amplitude of the measured ion velocity matches well with theoretical predictions over all $\omega_{d}$.  The quality factor of the COM mode is limited by the presence of dark ions produced by background gas collisions and by power-supply instabilities, but is high relative to the Fourier-limited linewidth of $\omega_{Z}$ for the values of $t_{d}$ shown here.  Agreement between data and theory breaks down for the largest values of $t_{d}$ which induce the largest ion velocities.  Under these conditions, the ions are driven out of the linear-response regime of Doppler velocimetry, or even to the blue side of the resonance, resulting in reduced scatter rates manifested as dips in the central oscillation lobe (Fig. 2j, l).  Reducing the RF excitation removes these effects.  The nonlinear velocimetry response can also be used as an independent calibration of the applied force (via Eq. 2 near resonance), agreeing to within a factor of order unity with the electric-field calibration described above (see \textit{Supplementary Information}).
\\
\indent We analyze the force detection sensitivity by systematically reducing the excitation strength and examining the system response through Fourier analysis.  Fig. 3a shows a one-dimensional slice of the time-domain Doppler velocimetry signal for diminishing values of $F_{d}$, accomplished by reducing $P_{RF}$ ($t_{d}$ is fixed at 1 ms) using a planar array of $n=$130$\pm$10 ions.  In order to minimize the number of required experimental cycles we employ a multichannel scaler and collect more than one photon ($\sim$3-5) per experimental cycle.  $F_{d}^{(ion)}$ is varied from $F_{0}^{(ion)}$ to $F_{0}^{(ion)}/100$, and the magnitude of the system response is reduced commensurately.  We Fourier transform these data (omitting the systematic delay for short times) and display the results on a semilog plot in Fig. 3b, demonstrating a spectral peak near the COM resonance frequency with diminishing signal-to-noise ratio (SNR).  
\\
\indent With $F_{d}^{(ion)}\equiv F_{0}^{(ion)}/100=$ 2.9$\pm$0.18 yN/ion and $n=$130$\pm$10, the total force on the array is $F_{d}=$ 377$\pm$37 yN.  From the spectral peak at $\omega_{Z}$ we calculate the SNR (here 2.3) and account for the measurement bandwidth to extract a force-detection sensitivity $\sim$1,200 $\textrm{yN}/\sqrt{\textrm{Hz}}$ for SNR $=1$ (see \textit{Supplementary Information}). 
\\
\indent The COM axial amplitude of an excitation with may be expressed as
\begin{equation}
z_{COM}=\frac{F_{d}t_{d}}{2nm_{Be}\omega_{Z}}.
\end{equation}
\noindent We characterize sensitivity to spatial displacement by considering small values of both $F_{d}$ and $t_{d}$.  For an excitation (as above) with $t_{d}=$1 ms and $F_{d}=$ 377 yN and $n=$130, we are able to discriminate spatial displacements of $z_{COM}\approx18$ nm (sensitivity $\sim58\;\textrm{nm}/\sqrt{\textrm{Hz}}$).  This amplitude is comparable to the thermal axial extent of the COM mode for a crystal of 130 ions (11 nm for $T=$ 0.5 mK), and an order of magnitude less than the thermal axial extent of an individual ion in the array (120 nm).  These absolute and relative sizes indicate that real-space imaging of ion excitation is not possible using, e.g. a CCD camera.  An excitation of this magnitude is also comparable to common values associated with nanomechanical resonators.  
\\
\indent Characterization of the minimum achievable force detection sensitivity is accomplished by producing a large motional response to a given applied force.  For a given $F_{d}^{(ion)}$ the Doppler velocimetry signal near resonance grows linearly with $t_{d}$, but detection sensitivity only increases (gets worse) as $\sqrt{t_{d}}$.  We have confirmed this scaling using ion crystals of $n=$ 60, 130, and 530 ions, observing an approximately linear increase in measured SNR up to $t_{d}=10$ ms (without bandwidth normalization).  For larger values of $t_{d}$ the experimental drifts of $\omega_{Z}$ on timescales comparable to the data acquisition time become significant relative to the narrowed Fourier response of the driven oscillator.
\\
\indent The smallest force we detected is $\sim$170~yN using a crystal of $n=60\pm5$ ions and $t_{d}=$10 ms, as described above.  Averaging over data acquired with different values of $F_{d}$ (normalizing by the SNR for different values of $F_{d}$ should yield similar values of force detection sensitivity) we find a minimum force detection sensitivity of 390$\pm$150 $\textrm{yN}/\sqrt{\textrm{Hz}}$.  The experimental uncertainty includes statistical fluctuations in averaging, uncertainty in the ion number, uncertainty in our calibration of the applied electric field, and imprecision in the calculation of the SNR.  The improvement in this value relative to that extracted from Fig. 3b is derived from both the increased drive time and the reduced ion number (see \textit{Supplementary Information}).  
\\
\indent It is important to distinguish between the measured systemic detection sensitivity (force or displacement) and the intrinsic sensitivity of the ions.  The ions respond to the external stimulus in a bandwidth set by $t_{d}^{-1}$, for these experiments $>$3000 times wider than $\tau_{M}^{-1}$.  Thus, the intrinsic ion sensitivity is at least 50 times better than reported in our measurement, and in line with sensitivities that may be derived from heating rate measurements~ \cite{Turchette2000, Deslauriers2004, Labaziewicz2008}.  However, in any experiment it is necessary to extract information from the system and account for inefficiencies in the readout technique when calculating the \emph{system's} sensitivity.  
\\
\indent Considerable systematic improvement is possible before our system becomes limited by the intrinsic sensitivity of trapped ions to forces and fields.  Realistic experimental improvements may be achieved by increasing the measurement bandwidth, primarily through improved light-collection efficiency (measurements are currently shot-noise limited).  For instance, experimental modifications allowing for large-solid-angle light collection and sampling during measurement at the Nyquist limit both reduce the required number of experimental cycles, $N$, and hence $\tau_{M}$.  Such improvements could allow realization of force-detection sensitivity $\sim$1.7 $\textrm{yN}/\sqrt{\textrm{Hz}}$, comparable to previously published calculations~\cite{Maiwald2009} (see \textit{Supplementary Information}).  Additionally, systematic improvements that increase the stability of $\omega_{Z}$, will permit longer drive times and hence improved force-detection sensitivities;  our system is currently limited by the effects of background gas collisions and power-supply instabilities.  
\\
\indent We have successfully employed this technique for the detection of optical dipole forces induced by Raman lasers.  Additionally, the Penning trap permits experiments using well over $n\approx10^{6}$, thus allowing optimization of sensitivity to electric \emph{fields}  (large total charge desirable) rather than applied \emph{forces} (small total mass desirable).  Given a crystal of this size and similar experimental conditions we could likely achieve an electric \emph{field} sensitivity of $\sim$500 $\textrm{nV}\; \textrm{m}^{-1}/\sqrt{\textrm{Hz}}$, with intrinsic sensitivity limits still lower.  
\\  
\indent The experiments we have presented have demonstrated phase-coherent excitation and detection of yN-level forces and nm-scale displacements induced by oscillating electric fields using trapped ions.  Our measurements have validated published calculations~\cite{Maiwald2009} suggesting that force-detection sensitivities of $\sim$1 $\textrm{yN}/\sqrt{\textrm{Hz}}$ are possible for single ion experiments, owing to the light mass of harmonically bound trapped ions and the presence of a strong readout technique.  Vitally, we have included in our detection-sensitivity measurements all relevant readout times, experimental dead times, hardware delays, and the like, making this a very conservative estimate of achievable force-detection sensitivity.  
\\
\indent Beyond simply measuring extremely small forces, the detection technique we employ provides an ability to discriminate motional excitations more than an order of magnitude smaller than the thermal extent of a single ion, and deep below the resolution-limits imposed by typical imaging systems.  Because the size of an ion's motional excursion scales as $\omega_{Z}^{-1}$, for forces of the magnitude we detect it is difficult to directly image motional excitations at frequencies larger than $\sim$0.1 MHz.  Thus phase-sensitive detection of ion motion may enable \emph{high-frequency} measurements as a part of a tunable, broadband detector.  
\\
\indent One potential shortcoming of this detection mechanism relates to the use of a phase-synchronous protocol.  In circumstances such as detector calibration or the application of an optical dipole force it is straightforward to synchronize detection events to the externally applied force.  In general the problem becomes significantly more complicated when trying to measure an unknown force.  However, it is typically possible to modulate an unknown external signal near $\omega_{Z}$ in such a way as to permit synchronization of detection events~\cite{HorowitzHill}.  For instance, in surface science or field mapping studies such as proposed in ~\cite{Maiwald2009}, a drive signal in the target device (e.g. a current bias) could be modulated directly, producing a time varying, detectable force or field.  Synchronization with the modulation envelope signal would then permit performance similar to that achieved in our experiments.  In such an approach the use of an ion-based detector could provide a means to map forces and fields at target-probe distances not generally accessible using cantilevers or other comparable approaches.  As an additional example, in studying the dynamics of surface charge accumulation, a laser producing photo-excited charge could be modulated in order to provide a synchronization signal~\cite{BlattField2010}. Doing so, and then using phase-synchronous detection, would permit a study of high-frequency charging dynamics that are not accessible in standard ``time-integrated'' detection techniques.
\\
\indent The results we have presented build on a strong foundation laid by previous trapped-ion experiments, and suggest that ion-based sensors may form a vital tool for sensitive measurements in nanoscale science.  Their utility will be maximized in applications where lateral spatial resolution may be traded against increased overall sensitivity and the ability to detect forces and fields with standoff distances greater than achievable using detectors based on integrated nanodevices.  Realistic, field-deployable sensors will likely require the use of integrated ion-trap chips with asymmetric potentials for directional force detection and three-dimensional spatial mapping.  It may also be desirable to exploit sub-Doppler cooling and sideband-detection mechanisms~\cite{Bible, Turchette2000} for the measurement of stochastic fields and forces.  
\\
\indent \textit{Acknowledgements} The authors thank K. Lehnert, D. Leibfried, D. J. Reilly, T. Rosenband, and D. J. Wineland for useful discussions.  We also thank J. Kitching and U. Warring for their comments on the manuscript.  We acknowledge research funding from DARPA and the NIST Quantum Information Program.  M.J.B. acknowledges fellowship support from IARPA and Georgia Tech, and H.U. acknowledges support from CSIR.  This manuscript is a contribution of NIST, a US Government Agency, and is not subject to copyright.
\\
\indent \textit{Author Contributions}  M.J.B., H.U., and J.J.B. conceived, designed, and performed the experiments and co-wrote the manuscript.  J.P.B. contributed to experimental measurements and data analysis.  A.P.VD. assisted with optics hardware and hardware-software interface. 
\\
\indent \textit{Additional Information} Supplementary information accompanies this paper at www.nature.com/naturenanotechnology. Reprints and permission information is available online at http://npg.nature.com/reprintsandpermissions/.  Correspondence and requests for materials should be addressed to M.J.B.
\\
\\

\begin{figure}[htbp]
\begin{center}
\includegraphics[width=8cm]{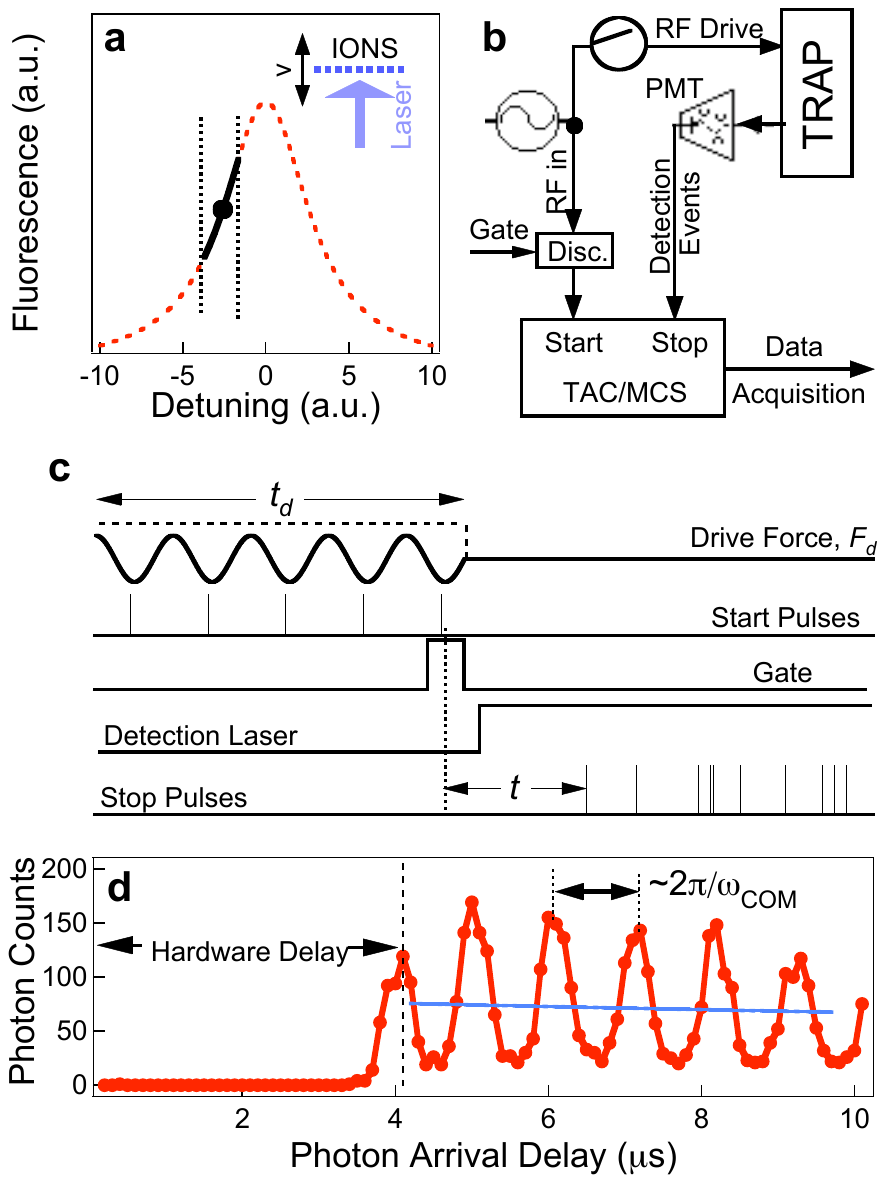}
\caption{\scriptsize Phase-coherent Doppler velocimetry.  (a) Schematic of the atomic resonance employed for detection of ion motion, $\gamma/2\pi\approx$19 MHz as a function of the detection laser detuning from the atomic resonance frequency.  The detection laser beam is oriented perpendicular to the plane of a 2D ion crystal (Inset).  An oscillating ion array periodically traces out a path on the atomic resonance profile schematically illustrated by the solid line.  The extent of the excursion is set by the magnitude of maximum ion velocity.  Vertical dashed lines indicate the range of Doppler shift associated with a given excursion along the resonance profile. (b) Schematic of the Doppler-detection system based on photon-arrival-time measurements.  Disc. = Discriminator,  TAC = Time-to-Amplitude Converter, MCS = Multi Channel Scaler,  PMT = photomultiplier tube.  (c) Schematic of pulse sequencing/triggering for phase-coherent detection.  This scheme is based on previous studies of micromotion nulling in a Paul trap~\cite{Berkeland1998} and studies of plasma oscillations in a Penning trap~\cite{mitchellDoppler}, with the important distinction that the excitation and detection are segregated into different parts of the measurement procedure.  $F_{d}=$ oscillating drive force due to the ac electric field, with dotted line showing the pulse envelope defining the pulsed-excitation period.  Start pulses synchronous with the Drive are output from a discriminator and fed into the TAC or MCS.  A Gate pulse ensures that only the last Start pulse of the excitation period triggers the TAC/MCS to begin data acquisition.  The Detection Laser is turned on using an RF switch controlling an Acousto-Optic Modulator, after which scattered photons may be detected.    Stop Pulses are generated by the detection of scattered photons using the PMT and are fed to the TAC/MCS.  (d) Histogram of photon arrival times relative to start-pulses generated synchronously with an RF drive of the COM mode on resonance.  Horizontal axis corresponds to the time delay, $t$ indicated in panel (c).  Photon arrivals are bunched with periodicity given by the driven COM oscillation period, following hardware delays.  Solid blue line is an exponential fit to the data used to remove a background scattering rate (see \textit{Supplementary Information}). }

\end{center}
\end{figure}

\begin{figure}[htbp]
\begin{center}
\includegraphics[width=8cm]{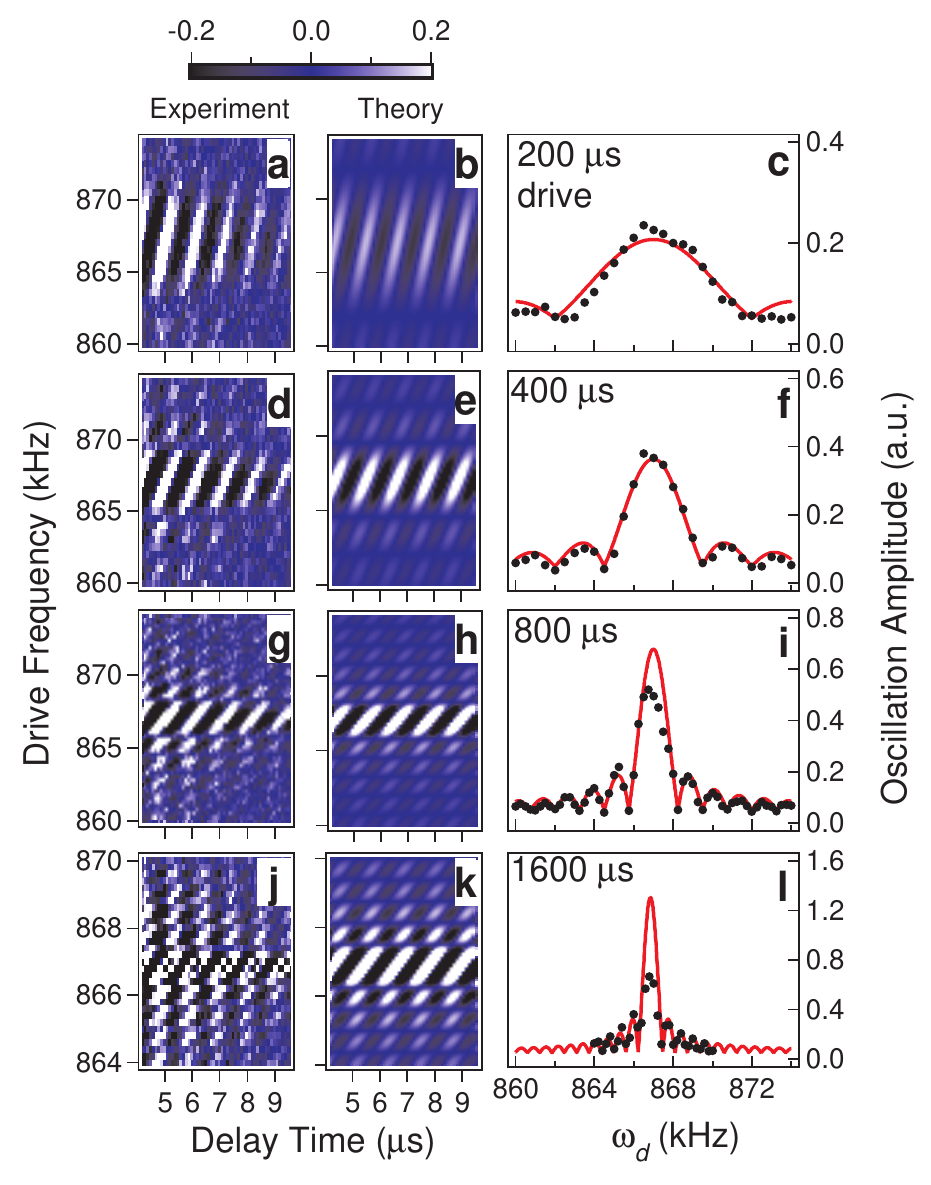}
\caption{\scriptsize Phase-coherent detection of the COM mode by RF excitation.  $F_{d}^{(ion)}=F_{0}^{(ion)}$.  First two columns: Residual of fit to exponential decay in photon arrival times for four driving pulse durations, expressed as a colorscale for experiment (a, d, g, j) and theory (b, e, h,  k).  Horizontal axis represents arrival delay from start pulse and vertical axis represents drive frequency.  Amplitude of oscillations decays with delay time due to radiation damping from the detection laser and is not accounted for in theoretical plots.  Third column: standard deviation of photon arrival times as a function of drive frequency.  Each data point represents the standard deviation of a horizontal slice of the two-dimensional plots (left) and illustrates resonant excitation of the COM mode as in Fig. 1c (standard deviation over the whole measurement period is used as a proxy for oscillation amplitude).  Solid lines represent theoretical fits using fixed drive times with force strength and a constant offset used as free parameters.  Each row of plots in the figure corresponds to a fixed drive time.  Fit parameters extracted from 200 $\mu$s drive duration used for longer drive periods.  Breakdown in fit quality on resonance increases with drive time as strong excitation leads the ions to be shifted out of the linear response regime to the blue side of the Doppler resonance, hence decreasing fluorescence.}

\end{center}
\end{figure}

\begin{figure}[htbp]
\begin{center}
\includegraphics[width=8cm]{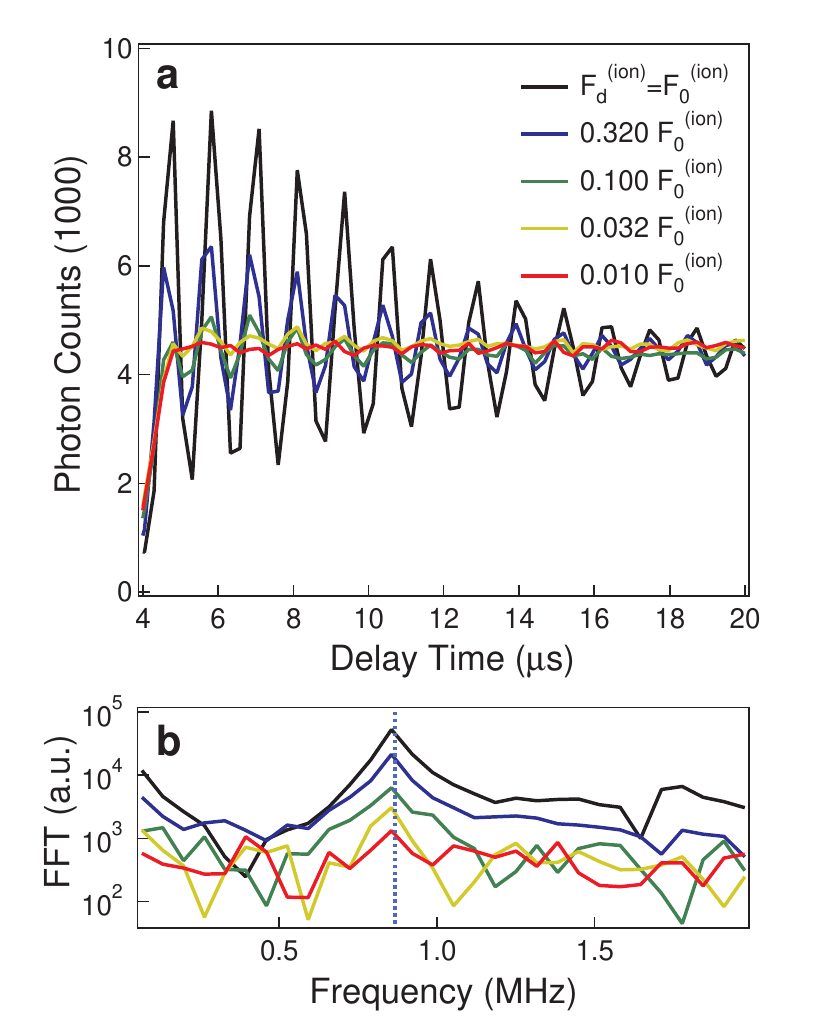}
\caption{\scriptsize Calibration of force-detection sensitivity by Fourier analysis.  (a) Temporal response to applied $F_{d}^{(tot)}=n F_{d}^{(ion)}$ with $n=130$ for decreasing drive strength and $t_{d}$ = 1 ms.  Total experiment time $\tau_{M}\approx$56 s for each trace including all measurement and dead time (recooling, photon detection, hardware delays), corresponding to $\sim$40,000 excitation/detection cycles.  Observed decay of oscillation magnitude due to radiation damping occurring once the detection laser is turned on.  (b) FFT of temporal response traces recorded in panel (a) above, plotted on a semilog scale, with same color-coding as in panel (a).  Spectral peak apparent at $\omega_{COM}$ diminishes in strength with decreasing $F_{d}^{(tot)}$.  Spectral peak has SNR $\approx$ 2.3 for $F_{d}^{(ion)}=0.010 F_{0}^{(ion)}$ (see text).}
\label{default}
\end{center}
\end{figure}

%\bibliography{npbib}

\begin{thebibliography}{10}
\expandafter\ifx\csname url\endcsname\relax
  \def\url#1{\texttt{#1}}\fi
\expandafter\ifx\csname urlprefix\endcsname\relax\def\urlprefix{URL }\fi
\providecommand{\bibinfo}[2]{#2}
\providecommand{\eprint}[2][]{\url{#2}}

\bibitem{Rugar2004}
\bibinfo{author}{Rugar, D.}, \bibinfo{author}{Budakian, R.},
  \bibinfo{author}{Mamin, H.~J.} \& \bibinfo{author}{Chui, B.~W.}
\newblock \bibinfo{title}{Single spin detection by magnetic resonance force
  microscopy}.
\newblock \emph{\bibinfo{journal}{Nature}} \textbf{\bibinfo{volume}{430}},
  \bibinfo{pages}{329--332} (\bibinfo{year}{2004}).

\bibitem{AFM1986}
\bibinfo{author}{Binnig, C.}, \bibinfo{author}{Quate, C.~F.} \&
  \bibinfo{author}{Gerber, C.}
\newblock \bibinfo{title}{Atomic force microscope}.
\newblock \emph{\bibinfo{journal}{Phys. Rev. Lett.}}
  \textbf{\bibinfo{volume}{56}}, \bibinfo{pages}{930--933}
  (\bibinfo{year}{1986}).

\bibitem{Casimir1998}
\bibinfo{author}{Mohideen, U.} \& \bibinfo{author}{Roy, A.}
\newblock \bibinfo{title}{Precision measurement of the casimir force from 0.1
  to 0.9 $\mu$m}.
\newblock \emph{\bibinfo{journal}{Appl. Phys. Lett.}}
  \textbf{\bibinfo{volume}{81}}, \bibinfo{pages}{4549--4552}
  (\bibinfo{year}{1998}).

\bibitem{Friction1999}
\bibinfo{author}{Volokitin, A.~I.} \& \bibinfo{author}{Persson, B. N.~J.}
\newblock \bibinfo{title}{Theory of friction: the contribution from a
  fluctuating electric field}.
\newblock \emph{\bibinfo{journal}{J. Phys. C}} \textbf{\bibinfo{volume}{11}},
  \bibinfo{pages}{345--359} (\bibinfo{year}{1999}).

\bibitem{Gravity1998}
\bibinfo{author}{Arkani-Hamed, N.}, \bibinfo{author}{Dimopoulos, S.} \&
  \bibinfo{author}{Dvali, G.}
\newblock \bibinfo{title}{The hierarchy problem and new dimensions at a
  millimeter}.
\newblock \emph{\bibinfo{journal}{Phys. Lett. B}}
  \textbf{\bibinfo{volume}{429}}, \bibinfo{pages}{263--272}
  (\bibinfo{year}{1998}).

\bibitem{Rugar2001}
\bibinfo{author}{Mamin, H.~J.} \& \bibinfo{author}{Rugar, D.}
\newblock \bibinfo{title}{Sub-attonewton force detection at millikelvin
  temperatures}.
\newblock \emph{\bibinfo{journal}{Appl. Phys. Lett.}}
  \textbf{\bibinfo{volume}{79}}, \bibinfo{pages}{3358--3360}
  (\bibinfo{year}{2001}).

\bibitem{Lehnert2009}
\bibinfo{author}{Teufel, J.~D.}, \bibinfo{author}{Donner, T.},
  \bibinfo{author}{Castellanos-Beltran, M.~A.}, \bibinfo{author}{Harlow, J.~W.}
  \& \bibinfo{author}{Lehnert, K.~W.}
\newblock \bibinfo{title}{Nanomechanical motion measured with an imprecision
  below that at the standard quantum limit}.
\newblock \emph{\bibinfo{journal}{Nature Nanotechnology}}
  \textbf{\bibinfo{volume}{4}}, \bibinfo{pages}{820--823}
  (\bibinfo{year}{2009}).

\bibitem{Cleland2003}
\bibinfo{author}{Knobel, R.~G.} \& \bibinfo{author}{Cleland, A.~N.}
\newblock \bibinfo{title}{Nanometer-scale displacement sensing using a
  single-electron transistor}.
\newblock \emph{\bibinfo{journal}{Nature}} \textbf{\bibinfo{volume}{424}},
  \bibinfo{pages}{291--293} (\bibinfo{year}{2003}).

\bibitem{Schwab2004}
\bibinfo{author}{Lahaye, M.~D.}, \bibinfo{author}{Buu, O.},
  \bibinfo{author}{Camarota, B.} \& \bibinfo{author}{Schwab, K.~C.}
\newblock \bibinfo{title}{Approaching the quantum limit of a nanomechanical
  resonator}.
\newblock \emph{\bibinfo{journal}{Science}} \textbf{\bibinfo{volume}{304}},
  \bibinfo{pages}{74--77} (\bibinfo{year}{2004}).

\bibitem{Lehnert2008}
\bibinfo{author}{Regal, C.~A.}, \bibinfo{author}{Teufel, J.~D.} \&
  \bibinfo{author}{Lehnert, K.~W.}
\newblock \bibinfo{title}{Measuring nanomechanical motion with a microwave
  cavity interferometer}.
\newblock \emph{\bibinfo{journal}{Nature Physics}}
  \textbf{\bibinfo{volume}{4}}, \bibinfo{pages}{555--560}
  (\bibinfo{year}{2008}).

\bibitem{brel88}
\bibinfo{author}{Brewer, L.~R.} \emph{et~al.}
\newblock \bibinfo{title}{Static properties of a non-neutral $^9$be$^+$ ion
  plasma}.
\newblock \emph{\bibinfo{journal}{Phys. Rev. A}} \textbf{\bibinfo{volume}{38}},
  \bibinfo{pages}{859--873} (\bibinfo{year}{1988}).

\bibitem{BiercukQIC2009}
\bibinfo{author}{Biercuk, M.~J.} \emph{et~al.}
\newblock \bibinfo{title}{High-fidelity quantum control using ion crystals in a
  penning trap}.
\newblock \emph{\bibinfo{journal}{Quantum Information and Computation}}
  \textbf{\bibinfo{volume}{9}}, \bibinfo{pages}{920--949}
  (\bibinfo{year}{2009}).

\bibitem{Heinzen1990}
\bibinfo{author}{Heinzen, D.~J.} \& \bibinfo{author}{Wineland, D.~J.}
\newblock \bibinfo{title}{Quantum-limited cooling and detection of
  radio-frequency oscillations by laser-cooled ions}.
\newblock \emph{\bibinfo{journal}{Physical Review A}}
  \textbf{\bibinfo{volume}{42}}, \bibinfo{pages}{2977--2994}
  (\bibinfo{year}{1990}).

\bibitem{Berkeland1998}
\bibinfo{author}{Berkeland, D.~J.}, \bibinfo{author}{Miller, J.~D.},
  \bibinfo{author}{Bergquist, J.~C.}, \bibinfo{author}{Itano, W.~M.} \&
  \bibinfo{author}{Wineland, D.~J.}
\newblock \bibinfo{title}{Minimization of ion micromotion in a paul trap}.
\newblock \emph{\bibinfo{journal}{J. Appl. Phys.}}
  \textbf{\bibinfo{volume}{83}}, \bibinfo{pages}{5025--5033}
  (\bibinfo{year}{1998}).

\bibitem{mitchellDoppler}
\bibinfo{author}{Mitchell, T.~B.}, \bibinfo{author}{Bollinger, J.~J.},
  \bibinfo{author}{Huang, X.-P.} \& \bibinfo{author}{Itano, W.~M.}
\newblock \bibinfo{title}{Doppler imaging of plasma modes in a penning trap}.
\newblock \emph{\bibinfo{journal}{Optics Express}}
  \textbf{\bibinfo{volume}{2}}, \bibinfo{pages}{314} (\bibinfo{year}{1998}).

\bibitem{Maiwald2009}
\bibinfo{author}{Maiwald, R.} \emph{et~al.}
\newblock \bibinfo{title}{Stylus ion trap for enhanced access and sensing}.
\newblock \emph{\bibinfo{journal}{Nature Physics}}
  \textbf{\bibinfo{volume}{5}}, \bibinfo{pages}{551--554}
  (\bibinfo{year}{2009}).

\bibitem{Bible}
\bibinfo{author}{Wineland, D.~J.} \emph{et~al.}
\newblock \bibinfo{title}{Experimental issues in coherent quantum-state
  manipulation of trapped atomic ions}.
\newblock \emph{\bibinfo{journal}{J. Res. NIST}}
  \textbf{\bibinfo{volume}{103}}, \bibinfo{pages}{259} (\bibinfo{year}{1998}).

\bibitem{Turchette2000}
\bibinfo{author}{Turchette, Q.~A.} \emph{et~al.}
\newblock \bibinfo{title}{Heating of trapped ions from the ground state}.
\newblock \emph{\bibinfo{journal}{Phys. Rev. A}} \textbf{\bibinfo{volume}{61}},
  \bibinfo{pages}{063418} (\bibinfo{year}{2000}).

\bibitem{Deslauriers2004}
\bibinfo{author}{Deslauriers, L.} \emph{et~al.}
\newblock \bibinfo{title}{Zero-point cooling and low heating of trapped
  $^{111}$cd$^{+}$ ions}.
\newblock \emph{\bibinfo{journal}{Phys. Rev. A}} \textbf{\bibinfo{volume}{70}},
  \bibinfo{pages}{043408} (\bibinfo{year}{2004}).

\bibitem{Labaziewicz2008}
\bibinfo{author}{Labaziewicz, J.} \emph{et~al.}
\newblock \bibinfo{title}{Temperature dependence of electric field noise above
  gold surfaces}.
\newblock \emph{\bibinfo{journal}{Phys. Rev. Lett.}}
  \textbf{\bibinfo{volume}{101}}, \bibinfo{pages}{180602}
  (\bibinfo{year}{2008}).

\bibitem{Drewsen2004}
\bibinfo{author}{Drewsen, M.}, \bibinfo{author}{Mortensen, A.},
  \bibinfo{author}{Martinussen, R.}, \bibinfo{author}{Staanum, P.} \&
  \bibinfo{author}{Sorensen, J.~L.}
\newblock \bibinfo{title}{Nondestructive identification of cold and extremely
  localized single molecular ions}.
\newblock \emph{\bibinfo{journal}{Phys. Rev. Lett.}}
  \textbf{\bibinfo{volume}{93}}, \bibinfo{pages}{243201}
  (\bibinfo{year}{2004}).

\bibitem{Staanum}
\bibinfo{author}{Staanum, P.~F.}, \bibinfo{author}{Hojbjerre, K.} \&
  \bibinfo{author}{Drewsen, M.}
\newblock vol. \bibinfo{volume}{5, Applications of Ion Trapping Devices} of
  \emph{\bibinfo{series}{Practical aspects of trapped ion mass spectrometry}},
  chap. \bibinfo{chapter}{Sympathetically-cooled single ion mass spectrometry}
  (\bibinfo{publisher}{CRC Press}, \bibinfo{address}{Boca Raton, FL},
  \bibinfo{year}{2010}).

\bibitem{Major2005}
\bibinfo{author}{Major, F.~G.}, \bibinfo{author}{Gheorghe, V.~N.} \&
  \bibinfo{author}{Werthe, G.}
\newblock \emph{\bibinfo{title}{Charged Particle Traps: Springer Series on
  Atomic, Optical, and Plasma Physics}} (\bibinfo{publisher}{Springer},
  \bibinfo{address}{Berlin, Heidelberg, Germany}, \bibinfo{year}{2005}).

\bibitem{mitchell98}
\bibinfo{author}{Mitchell, T.~B.} \emph{et~al.}
\newblock \bibinfo{title}{Direct observations of structural phase transitions
  in planar crystallized ion plasmas}.
\newblock \emph{\bibinfo{journal}{Science}} \textbf{\bibinfo{volume}{282}},
  \bibinfo{pages}{1290} (\bibinfo{year}{1998}).

\bibitem{huap98a}
\bibinfo{author}{Huang, X.-P.}, \bibinfo{author}{Bollinger, J.~J.},
  \bibinfo{author}{Mitchell, T.~B.} \& \bibinfo{author}{Itano, W.~M.}
\newblock \bibinfo{title}{Phase-locked rotation of crystallized non-neutral
  plasmas by rotating electric fields}.
\newblock \emph{\bibinfo{journal}{Phys. Rev. Lett.}}
  \textbf{\bibinfo{volume}{80}}, \bibinfo{pages}{73--76}
  (\bibinfo{year}{1998}).

\bibitem{jenm05}
\bibinfo{author}{Jensen, M.~J.}, \bibinfo{author}{Hasegawa, T.},
  \bibinfo{author}{Bollinger, J.~J.} \& \bibinfo{author}{Dubin, D. H.~E.}
\newblock \bibinfo{title}{Rapid heating of a strongly coupled plasma near the
  solid-liquid phase transition}.
\newblock \emph{\bibinfo{journal}{Phys. Rev. Lett.}}
  \textbf{\bibinfo{volume}{94}}, \bibinfo{pages}{025001}
  (\bibinfo{year}{2005}).

\bibitem{HorowitzHill}
\bibinfo{author}{Horowitz, P.} \& \bibinfo{author}{Hill, W.}
\newblock \emph{\bibinfo{title}{The Art of Electronics}}
  (\bibinfo{publisher}{Cambridge University Press}, \bibinfo{address}{New York,
  NY}, \bibinfo{year}{2006}).

\bibitem{Rugar1997}
\bibinfo{author}{Stowe, T.} \emph{et~al.}
\newblock \bibinfo{title}{Attonewton force detection using ultrathin silicon
  cantilevers}.
\newblock \emph{\bibinfo{journal}{Appl. Phys. Lett.}}
  \textbf{\bibinfo{volume}{71}}, \bibinfo{pages}{288--290}
  (\bibinfo{year}{1997}).

\bibitem{Leibfried2003}
\bibinfo{author}{Leibfried, D.} \emph{et~al.}
\newblock \bibinfo{title}{Experimental demonstration of a robust high-fidelity
  geometric two ion-qubit phase gate}.
\newblock \emph{\bibinfo{journal}{Nature}} \textbf{\bibinfo{volume}{422}},
  \bibinfo{pages}{412--415} (\bibinfo{year}{2003}).

\bibitem{BlattField2010}
\bibinfo{author}{Harlander, M.}, \bibinfo{author}{Brownnutt, M.},
  \bibinfo{author}{Hansel, W.} \& \bibinfo{author}{Blatt, R.}
\newblock \bibinfo{title}{Trapped-ion probing of light-induced charging effects
  on dielectrics}.
\newblock \emph{\bibinfo{journal}{arxiv.org}} \bibinfo{pages}{1004.4842}
  (\bibinfo{year}{2010}).

\end{thebibliography}

\end{document}